\documentclass{sig-alternate}

\usepackage{comment}
\usepackage{color}
\usepackage{textcomp}
\usepackage{multirow}

\usepackage[utf8]{inputenc}
\usepackage[T1]{fontenc}

\begin{document}
%

\title{Exploring EEG for Object Detection and Retrieval}

%
%

%
%
%
%

\numberofauthors{6} 
%
%

\author{Eva Mohedano$^1$, Amaia Salvador$^2$, Sergi Porta$^2$, Xavier Gir{\'o}-i-Nieto$^2$,\\ 
Graham Healy$^1$, Kevin McGuinness$^1$, Noel O'Connor$^1$ and Alan F. Smeaton$^1$\\
\\
$^1$Dublin City University, Dublin, Ireland \\ 
$^2$Universitat Politècnica de Catalunya, Barcelona, Catalonia/Spain
}



\maketitle
\begin{abstract}
This paper explores the potential for using Brain Computer Interfaces (BCI) as a relevance feedback mechanism in content-based image retrieval. We investigate if it is possible to capture useful EEG signals to detect if relevant objects are present in a dataset of realistic and complex images.  We perform several experiments using a rapid serial visual presentation (RSVP) of images at different rates (5Hz and 10Hz) on 8 users with different degrees of familiarization with BCI and the dataset. We then use the feedback from the BCI and mouse-based interfaces to retrieve localized objects in a subset of TRECVid images. We show that it is indeed possible to detect such objects in complex images and, also, that users with previous knowledge on the dataset or experience with the RSVP outperform others. When the users have limited time to annotate the images (100 seconds in our experiments) both interfaces are comparable in performance. Comparing our best users in a retrieval task, we found that EEG-based relevance feedback outperforms mouse-based feedback. The realistic and complex image dataset differentiates our work from previous studies on EEG for image retrieval.
\end{abstract}




\section{Introduction}
\label{sec:intro}

The exponential growth of visual content and its huge diversity has motivated considerable research on how documents can be retrieved according to user intentions when formulating a query. 
Advances in image processing and computer vision have provided tools for a perceptual and semantic interpretation of both the query and the indexed content. 
These tools have allowed the development of retrieval systems capable of processing queries by example and concepts.

Such retrieval systems are, however, often of limited utility for two reasons. First, computers are still not yet on par with human performance when it comes to a semantic annotation of visual content, because the mechanisms that human brain uses for visual processing are still not completely understood. Second, and most important, because queries are often too simple to completely describe user intent: both visual and textual queries often contain insufficient information for the retrieval system to correctly understand the true information need.

For this reason, the role of a human user during visual retrieval is critical, and his judgment about the correctness of the retrieved results can greatly speed up the search processes.
This kind of relevance feedback has been demonstrated to significantly improve retrieval performance in image \cite{Tong:2001:SVM:500141.500159, zhou2003relevance} and video \cite{Yan:2003:NPF:957013.957087,Amir:2005:MRF:1101826.1101832} retrieval.
Indeed, the potential for improvement with relevance judgements is arguably even more significant in visual information retrieval. For example, in object search type tasks, images and video clips are typically represented by high-dimensional feature vectors, and the retrieval task can be posed as one of learning to rank the images in the dataset given a small number of examples. With such high-dimensional representations and very limited examples, attempting to train complex models with large numbers of parameters can result in serious under- or overfitting. One possible workaround to this is to use a very simple model to retrieve initial results, and then use relevance feedback to gather more training examples prior to training a more complex ranking function.

Manually annotating images using a mouse, especially in a visual retrieval context, can be tedious and mentally exhausting. EEG-based brain computer interfaces offer a potential solution. Although EEG hardware is currently cumbersome to setup, one can reasonably expect that this will be improved in the future. Moreover, research has firmly established through rapid-serial visual presentation (RSVP) experiments, that the human brain’s response to a recognition stimulus produces well-defined, characteristic, and detectable response in the corresponding EEG \cite{luck2014introduction}, even when the user is not fully conscious of having seen the item of interest. Future EEG technology thus has the potential to facilitate a user providing ``hands-free'' relevance feedback in a natural and un-intrusive way.

This paper explores this potential application by examining if it is indeed possible to reliably obtain reasonably accurate feedback on image relevance from a user using brain computer interfaces, and how such feedback compares with manual feedback provided by a mouse. We carry out several experiments designed to explore the potential for EEG as a mechanism for relevance feedback in object search tasks. First, we examine if it is indeed possible to obtain a useful relevance signal from EEG  on realistic instance search tasks (from TRECVid benchmark) chosen specifically to be very challenging. Second, we assess the importance of familiarity with the dataset on performance by contrasting accuracy between expert and non-expert users. Third, we evaluate different RSVP rates to determine which is most effective in terms of accuracy and speed. Finally, we directly compare the EEG-based relevance feedback signal with standard ``click-based'' relevance feedback from a mouse and find that comparable accuracy can be achieved with the EEG-based approach, especially for those users who are familiar with both the brain computer interface and the image dataset.

The remainder of the paper is organized as follows. Section~\ref{sec:relatedwork} reviews previous work exploring the use of EEG signals for multimedia analysis. Section~\ref{sec:methodology} describes the adopted methodology for our experiments. Section \ref{sec:imageclassification} describes the experiments using a BCI device and the impact of different parameters and configurations. Section~\ref{sec:eegvsmouse} compares the EEG-based annotation with the traditional mouse-based one. Section~\ref{sec:RelevanceFeedback} uses the feedback obtained from the collected annotations to retrieve images in a large dataset. Finally, Section~\ref{sec:discussion} draws conclusions and gives insights on future lines of research.
\section{Related Work}
\label{sec:relatedwork}

Previous work combining BCI and computer vision~\cite{bigdely2008brain, wang2009brain, Mohedano:2014:OSI:2647868.2654896} have been focused primarily on image classification and object detection, by annotating all the images in a collection. 



EEG signals have been used in \cite{bigdely2008brain} to detect airplanes in a dataset of satellite images from the city of London. In this case the visual inspection guided by RSVP is made expanding a small region of the image and reducing the contrast of the rest of the image. This preprocessing of the image guarantees the focus of the user's attention on the highlighted region. Although our work also aims at finding images containing certain instances of certain query objects, the appearance of both the object and their context present a higher diversity.

The work in \cite{healy2011optimising} expands the catalog of objects to detect, but in this case they are presented in very simple images where the object in a black background occupy the whole image. In our work we address a more complex scenario, where object instances are not central in the scene and appear in a complex context together with other people and objects. That work also used EEG to enhance the detection results obtained by asking users to push a button after every time they saw the target object in the RSVP. Our set up, while also asking users to push the button, completely ignores the button and drives its results from the EEG signals only.


The work in \cite{yang2009general} considers the application of relevance annotation to address the problem of inter-user variability when training a relevance SVM model on EEG signals. While the goal of the work was adapting the classifier of one user to another, the experiments were actually performed on complex keyframes from TRECVid 2005. The same authors ranked 4th that year in TRECVid interactive runs with a push-button RSVP interface with adjustable speed \cite{Hauptmann2006rsvp}, a result achieved with a user who was specifically trained for this task. Our experiments use similar images as in those works and report results for users with a diverse degree of experience, mainly novice. However, while that work aimed at detecting concepts depicted by the whole image, we focus on the more challenging task of detecting a local object in a complex scenario.

Retrieval with EEG was explored in \cite{wang2009brain} by formulating it as a semi-supervised learning problem. The noisy relevance labels generated by EEG classifiers on a small subset of a large database are expanded through a visual similarity graph. This graph is built by creating a vertex for each image in the database and establishing links between them with weights according to the visual similarity between images. Images are generally more complex than in \cite{healy2011optimising} as in most cases objects appear in a real context, but the object is still the main focus of the photo and contexts tend to be plain. In our experiments, the highest and lowest EEG relevance scores are used to train a binary classifier, instead of feeding a semi-supervised algorithm. This way results are more comparable to the binary relevance labels collected with the mouse interface.

EEG signals were replaced by much more precise fMRI-derived brain responses in \cite{han2013representing} in a retrieval scenario also based on TRECVid 2005 data  at a global scale.
In this case brain activity is analyzed to predict the semantics of the video which is being watched by the user.
On the other hand, a mapping of computer-generated visual features to semantic features is also learned, allowing this way the search in a visually indexed dataset based on a fMRI-derived query.
These features were compared to a classic BoW features and the reported experiments show that using brain-based semantic features performed better than retrieving based on the visual BoW features.
The main drawback of fMRI signals is its much higher cost when compared to EEG signals considered in our work.

The problem posed by the noisy EEG labels provided by the relevance scores provided by the classifiers was also addressed in a much more local scale in \cite{Mohedano:2014:OSI:2647868.2654896}.
That work aimed at segmenting a salient object from an image by building a RSVP of local areas of the image.
The noisy saliency map generated by assigning the EEG score of each window to its pixels was later smoothed with a Gaussian filtered and finally binarized to feed a segmentation algorithm.
This binarization step is similar to the once we apply but, in our case, not all the images (windows) are annotated, so we actually expand the binarized noisy labels to the unseen images in the database.


\section{Methodology}
\label{sec:methodology}

This section explains the set-up used in our experiments. First, we introduce the procedure to acquire the data from users using a BCI and the image dataset that we have used. Second, we introduce the used pipeline to process the EEG signals and use them to classify images.

\subsection{Data acquisition}
\label{sec:dataAcquisition}
\subsubsection{EEG signature: P300}
\label{ssec:EEGsignature}
Based on previous work on image retrieval based on EEG \cite{healy2011optimising,huang2011framework,wang2009brain}, the experimental design was based in the `oddball paradigm,' in which two different stimulus are presented to the user in random order and with different probabilities. One stimulus (the `target') appears with low probability during the visual presentation and the other (the `distractor') appears frequently. Users are asked to focus on detecting the target stimulus and to express their reaction by either counting or pressing a button.

In this context, it has been found that when a user reacts to a target stimulus, a P300 wave appears in the captured brain signals. This wave is a kind of Event Related Potential (ERP) and it consists in a positive peak in the EEG activity around the 250ms-500ms range after the stimulus presentation. Its amplitude and latency can vary across users and depend on many factors such as the improbability and difficulty of the targets in the stimulus presentation.

\begin{figure}[!ht]
  \begin{center}
    \includegraphics[width=\columnwidth]{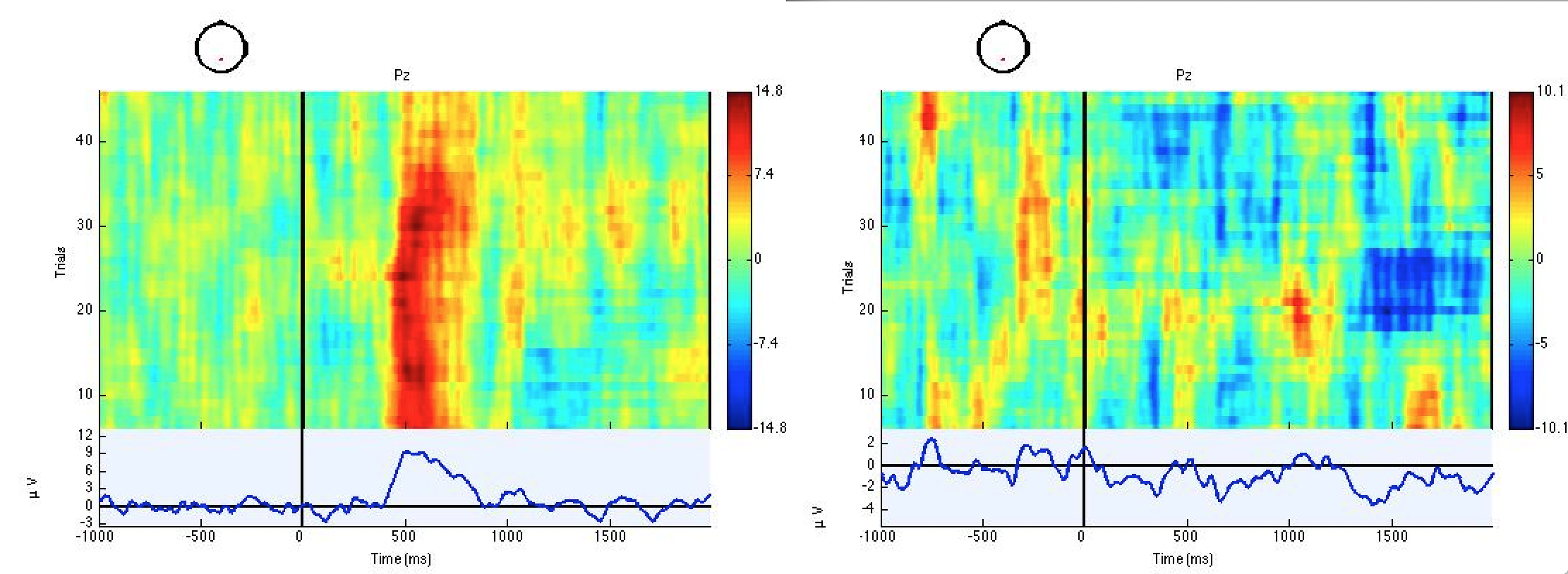}
    \caption[]{Visual evidence of discriminative P300 response on averaged epochs for 1000 images (5 blocks of 200) captured at 5Hz for the Pz channel. On the left, the response on target images. On the right, the response for 50 distractors.}
    \label{fig:p300}
  \end{center}
\end{figure}

\subsubsection{Dataset}
\label{sec:dataset}

\begin{figure}[h!]
    \centering
    \includegraphics[width=\columnwidth]{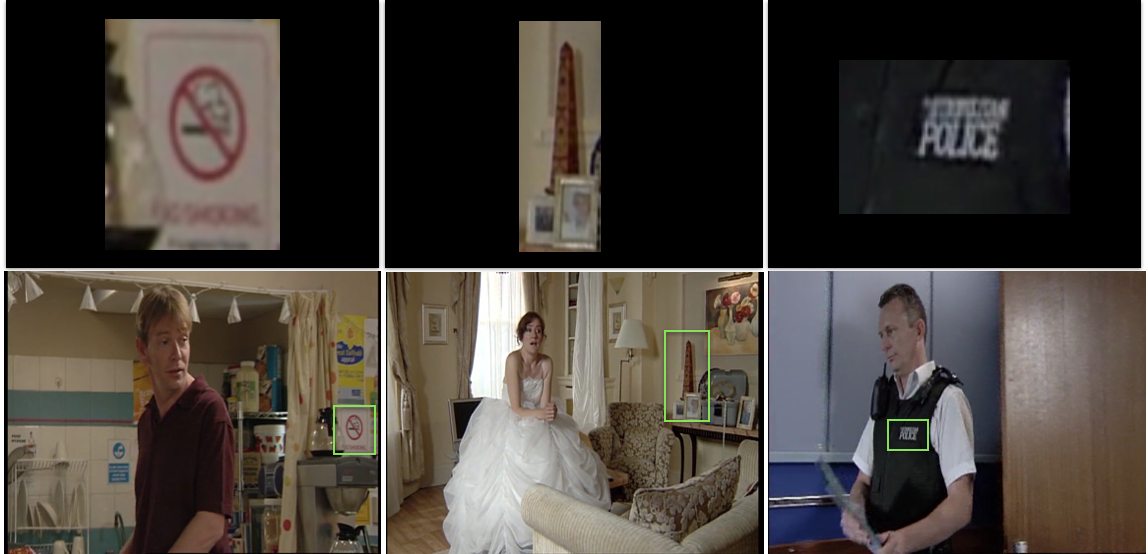}
    \caption{Visual examples of the three query objects (top) and the actual images where they appear (bottom). From left to right: 1. A circular `no smoking' logo, 2. A small red obelisk, 3. A Metropolitan Police logo.}
    \label{fig:queries}
\end{figure}

We selected a `real word' set of images from a subset of the TRECVid 2013 instance search dataset. Three queries were selected for the experiment, each one containing 4 visual examples. The selected queries contain small objects that appear in complex images. Figure~\ref{fig:queries} shows one visual example of each of the three selected queries. Using the ground truth labels provided by TRECVid, 1000 images were selected for each query. To adapt the data to the oddball paradigm, the `target' ratio of the dataset has been set to 5\% (i.e. for each query, 50 relevant images and 950 non-relevant images are used). 




  
\subsubsection{Image presentation and user task}
\label{ssec:RSVP}

The selected images are presented in a Rapid Serial Visual Presentation. For each query, the 1000 images are divided in 5 blocks of 200 images keeping a 5\% target ratio within each block (i.e. 10 `target' images and 190 `distractors'). Once this division is done, the images are randomly sorted within the block. These blocks are presented to the user one after the other, allowing him/her to rest for a few seconds between blocks. The visualization rates that have been tested in our experiments are 5Hz and 10Hz, therefore the duration of the experiment for a single query is 200 and 100 seconds.

The procedure to capture brain signals for a single query goes as follows: First, a textual description of the object of interest is given, followed by 4 images in which the object appears. After that, the 5 blocks are presented to the user through RSVP. The experiment is repeated on the user for the three queries, giving him/her some time to rest at the end of each presentation. In all the experiments that we performed, users were asked to press a button every time that they detected a target image.

Eight volunteers between the ages of 19 and 33 participated in the experiments: six women and two men. Of the eight users that participated in the task, two completed the experiment with a RSVP at 10Hz and four at 5Hz. The last two users did the task twice, once at 10Hz and once at 5Hz. 

The BCI device used was a 32 channels actiCHamp amplifier, which was connected to the user locating the electrodes according to the 10-20 system.




   

\subsection{EEG signal processing}
\label{sec:eegprocessing}

The following describes the procedure used to clean and classify the brain signals that we used to obtain EEG relevance scores for the images in the dataset.

\subsubsection{Feature vectors}

The signals captured by each one of the 32 sensors were re-referenced to the average of all the channels. Then, the sample rate was reduced from the original 1000Hz to 250Hz and the signals were  band-pass filtered from 0.1Hz to 20Hz. The epochs related to each visual stimulus were extracted, obtaining 1000 epochs with EEG activity from 1 second before to 2 seconds after each image. At this stage each image had 32 signals of 750 samples associated. Then, for each of the 32 signals, the period from 200ms to 1s was taken as the discriminant time region to discern between EEG responses of targets and distractors (see Figure~\ref{fig:p300}). We select the activity from 200ms to 1s after the stimulus presentation and we reduce the signal's sample rate from 250Hz to 20Hz, generating a 16 sample vector per channel. Each of the 16 samples per channel was the result of computing the average of 24 samples windows with 50\% of overlap between each other, which we found was a better strategy than to just decimate the signal. Finally, we build a single feature vector for the image as the concatenation of the 32 channels, generating a 512-dimension vector per image.

\subsubsection{Classification and EEG scores}

We used a linear SVM model with default parameters from the scikit-learn Python library to classify the EEG signals. Every user had his/her own models, since both the P300 signals and the noise associated to the EEG recordings can vary quite a lot across users and, for this reason, mixing data from different users can cause significant drops in the classifier performance. Thus, for each user, the data associated to the presentation of 2 queries (2000 EEG epochs) were used as training examples, and the remaining query was used for testing. This procedure was repeated 3 times to obtain a classification prediction for the images of the three queries.

We used the Area Under the Curve (AUC) of the Receiver Operating Characteristic space (ROC) to evaluate the performance of the models.

\section{Object detection}
\label{sec:imageclassification}

In this section we explore the potential of EEG signals to classify between relevant and non-relevant images. We also conduct a study trying to quantize the impact of user profiles and visualization rates in the system's accuracy. 

\subsection{Qualitative results}
\label{ssec:qualitativeresults}

\begin{figure}[!ht]
  \begin{center}
  	\includegraphics[width=0.3\columnwidth]{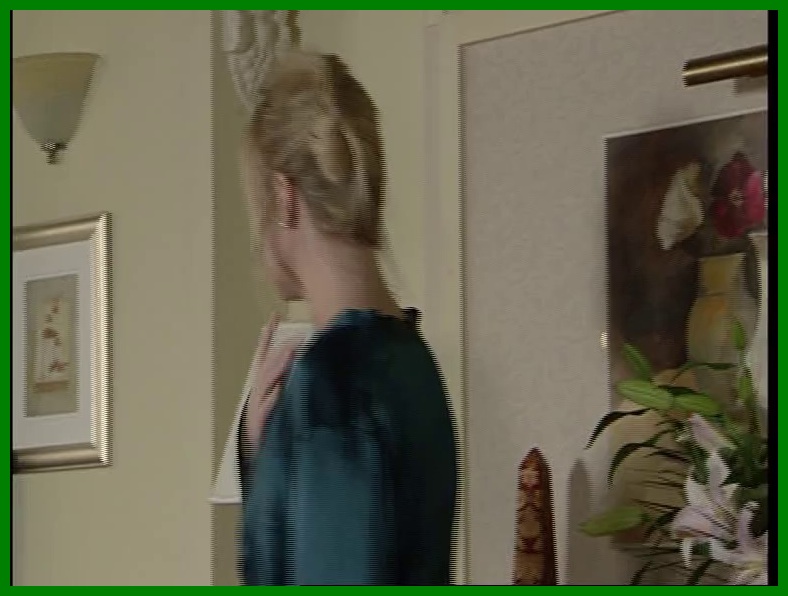}
    \includegraphics[width=0.3\columnwidth]{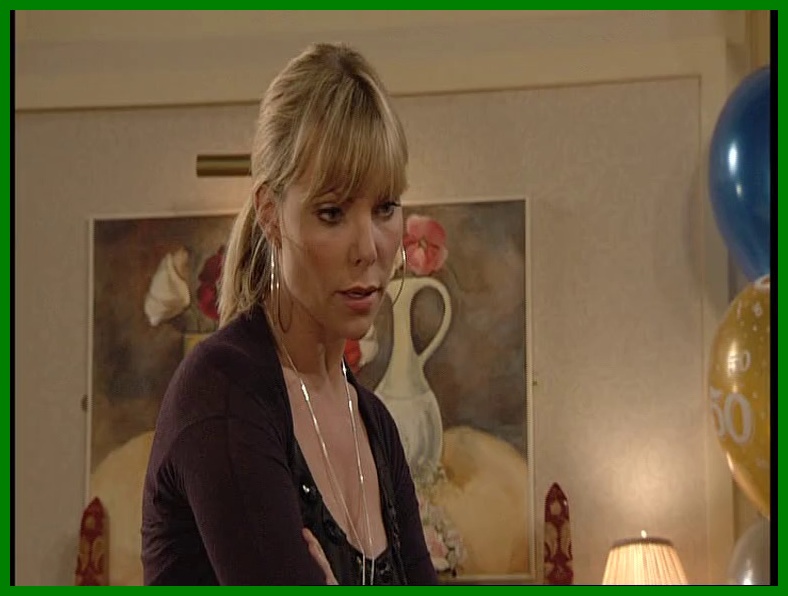}
    \includegraphics[width=0.3\columnwidth]{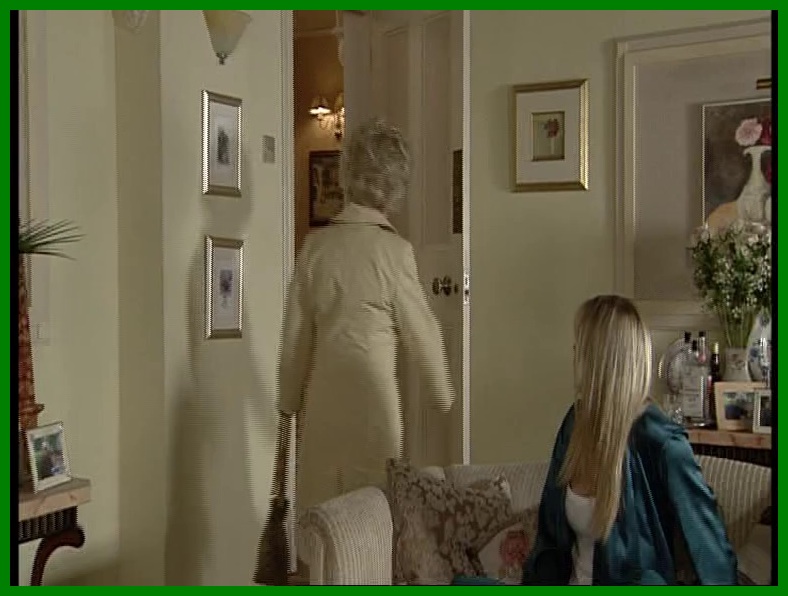}
  	\includegraphics[width=0.3\columnwidth]{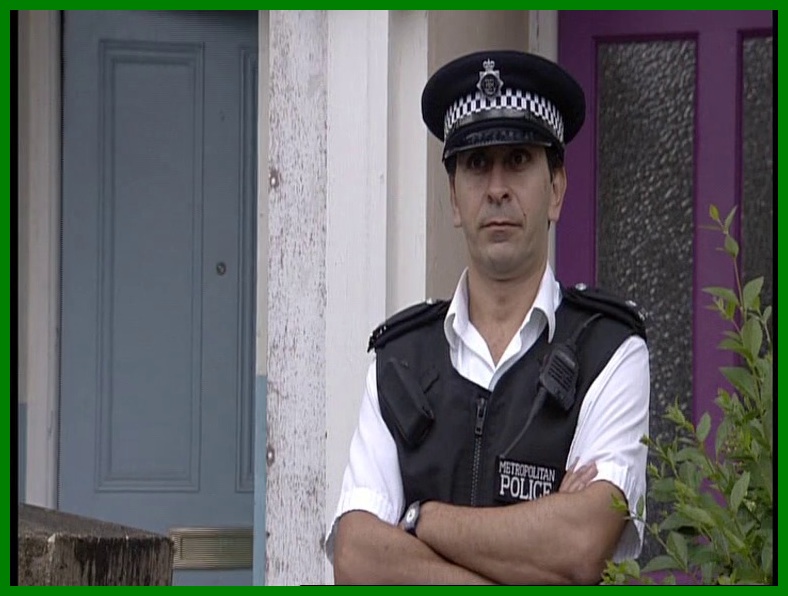}
    \includegraphics[width=0.3\columnwidth]{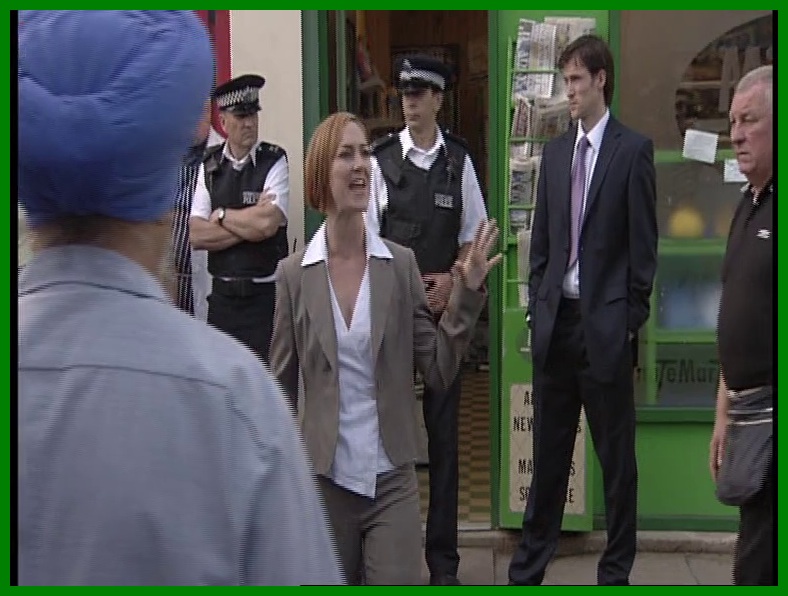}
    \includegraphics[width=0.3\columnwidth]{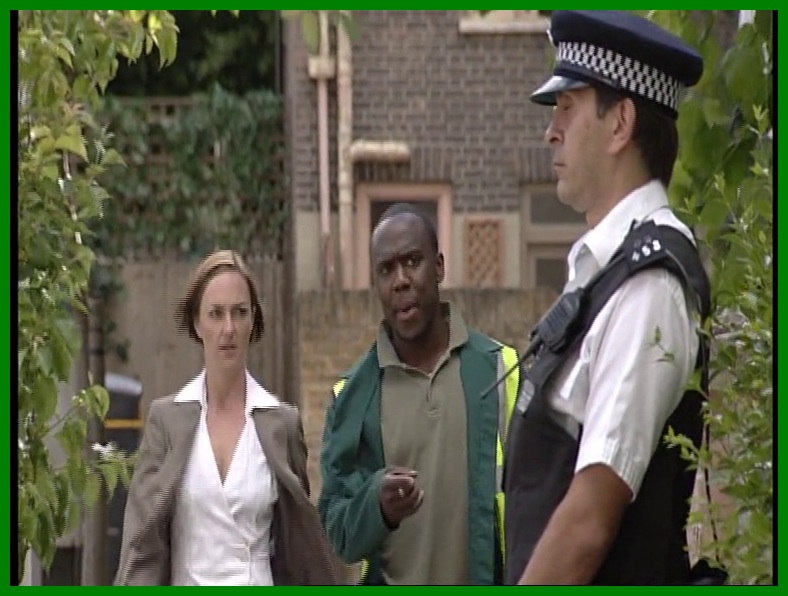}

    \caption[]{Images with high SVM classifier scores for user A and query 2 (top) and 3 (bottom). }
    \label{fig:relevant}
  \end{center}
\end{figure}

The three selected queries for these experiments refer to small objects that appear in complex images. One could think that these objects should be very difficult for users to recognize when images are presented at such high frequency. Figure~\ref{fig:relevant} shows a few examples of some of the correctly classified images for queries 2 and 3 (the red obelisk and the police logo). In these examples we can see images that have been correctly detected as relevant despite of the fact that, in most of them, the object appears to be very small (sometimes even incomplete). At the beginning of the experiment, users are shown 4 image examples where the object of interest appears, giving them the opportunity to learn about its context as well. The examples in Figure~\ref{fig:relevant} suggest that users do not only recognize the object itself, but also its context. For query 2 (the red obelisk), users might be recognizing the room in which the object is usually located, or the people who are usually around it, while for query 3 (the police logo), users might respond to images where there is a policeman, or someone wearing a dark vest.

\begin{figure}[!ht]
  \begin{center}
  	\includegraphics[width=0.3\columnwidth]{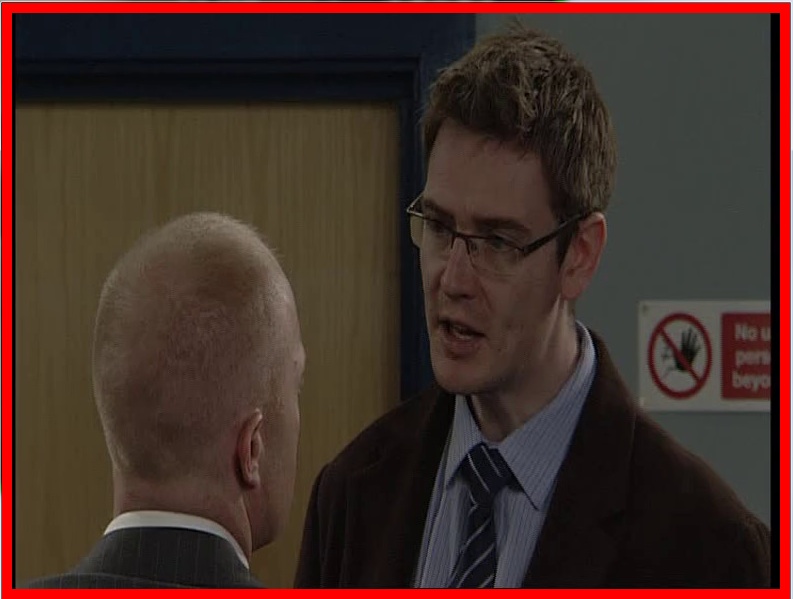}
    \includegraphics[width=0.3\columnwidth]{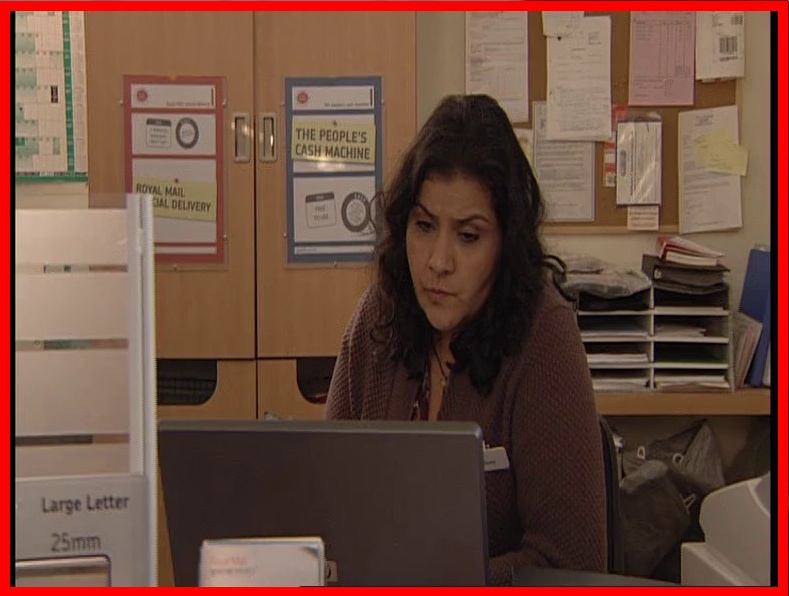}
    \includegraphics[width=0.3\columnwidth]{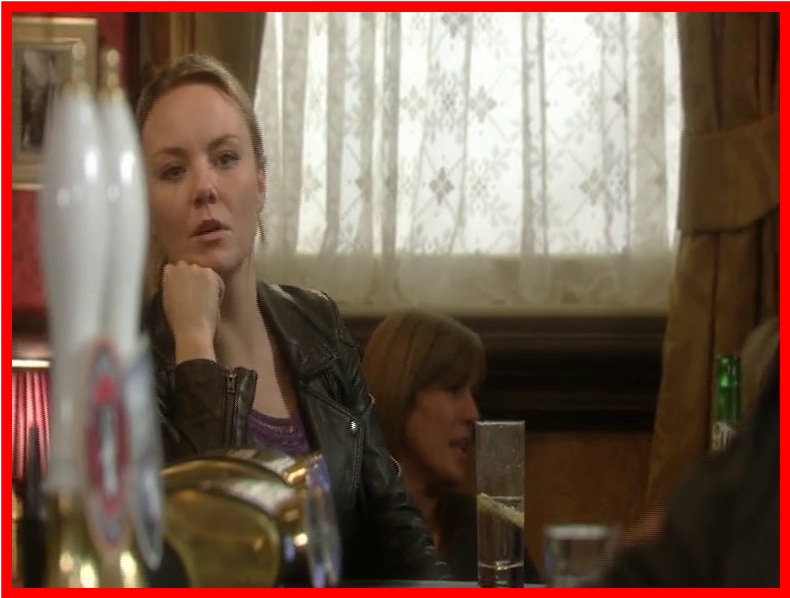}
    \includegraphics[width=0.3\columnwidth]{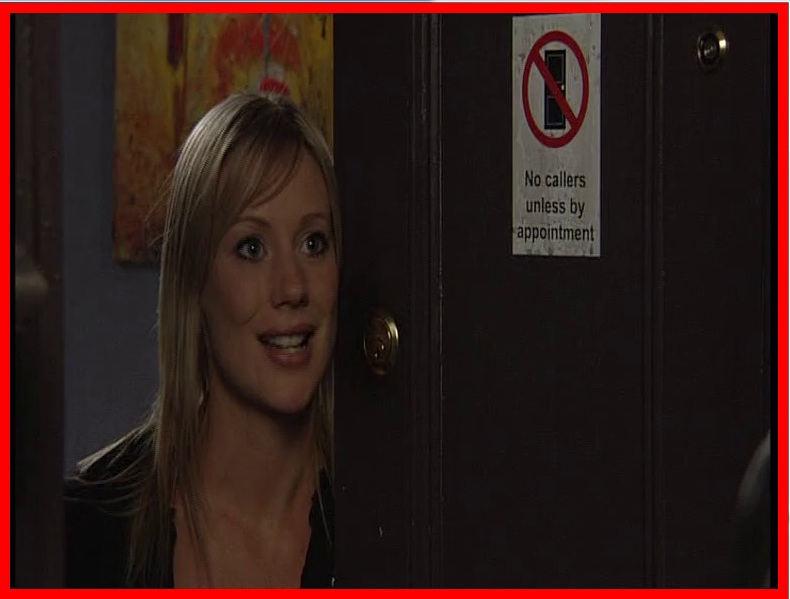}
    \includegraphics[width=0.3\columnwidth]{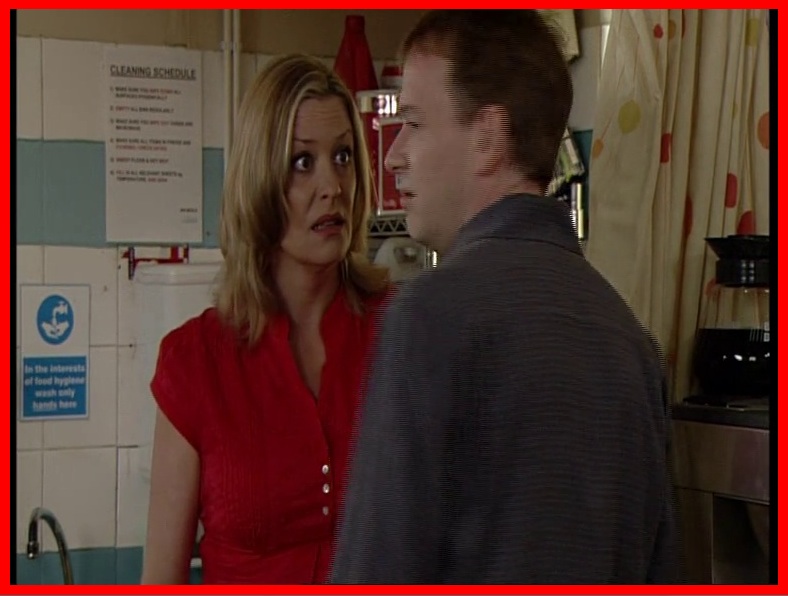}    
    \includegraphics[width=0.3\columnwidth]{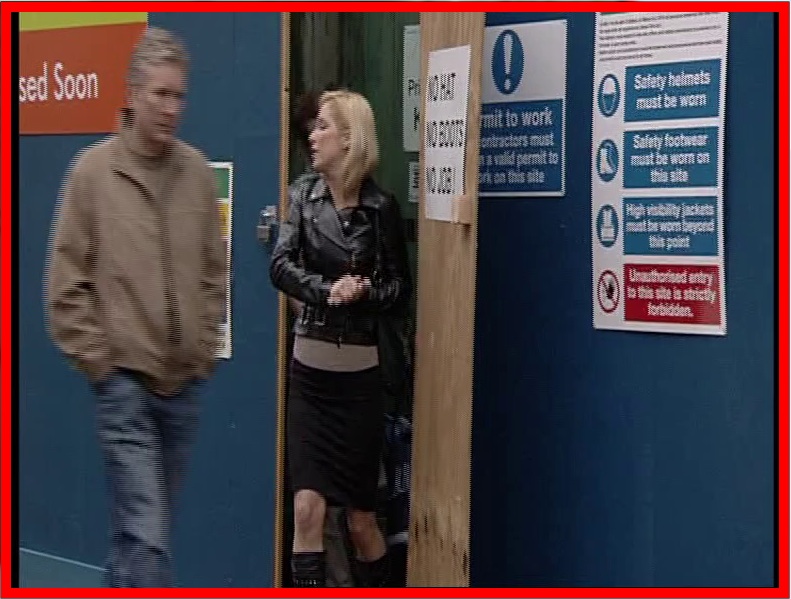}

    \caption[]{Images with high SVM classifier scores (yet not relevant) for user D and query 1. }
    \label{fig:mistakes}
  \end{center}
\end{figure}

From the examples in Figure~\ref{fig:relevant} one could think that users only respond to the general appearance of the image and that they are not really identifying the object of interest, but the type of mistakes that users make prove otherwise. Figure~\ref{fig:mistakes} shows some sample images that obtained high SVM scores but are actually not relevant to the query. It is interesting to see that all these images contain objects that are similar to the queried object: other logos, red round-shaped objects, etc. This also proves that users respond not only to the general appearance of the image but also to the object they are looking for.

\subsection{User diversity }
\label{ssec:users}

\begin{figure}[!ht]
  \begin{center}
    \includegraphics[width=\columnwidth]{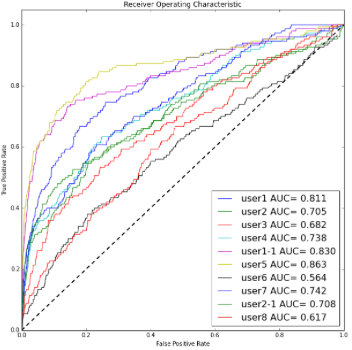}
    \caption[]{ROC curves for the 8 users. Users have as many curves as the number of experiments they performed.}
    \label{fig:roc}
  \end{center}
\end{figure}

Figure~\ref{fig:roc} shows the ROC curves for all the users who participated in the experiment. The first thing to notice is the huge diversity in users' performance. Such diversity could be explained by many factors (e.g. level of attention, fatigue, noise, etc.), but one thing that could also have an impact in the performance is the level of expertise of the user. Following this idea, we define two user types: 
\begin{itemize}
\item The \emph{expert} user, who is familiar with the presented images and with the purposes of the experiment. Those users who are directly related to this project have been considered experts, since they have spent a considerable amount of time looking at the images in the dataset. 
\item The \emph{novice} user, who has had no previous exposure to the images.
\end{itemize}

Following the criteria stated above, 3 of our users were selected as experts and 5 of them as novice users. 

\begin{figure}[!ht]
  \begin{center}
    \includegraphics[width=\columnwidth]{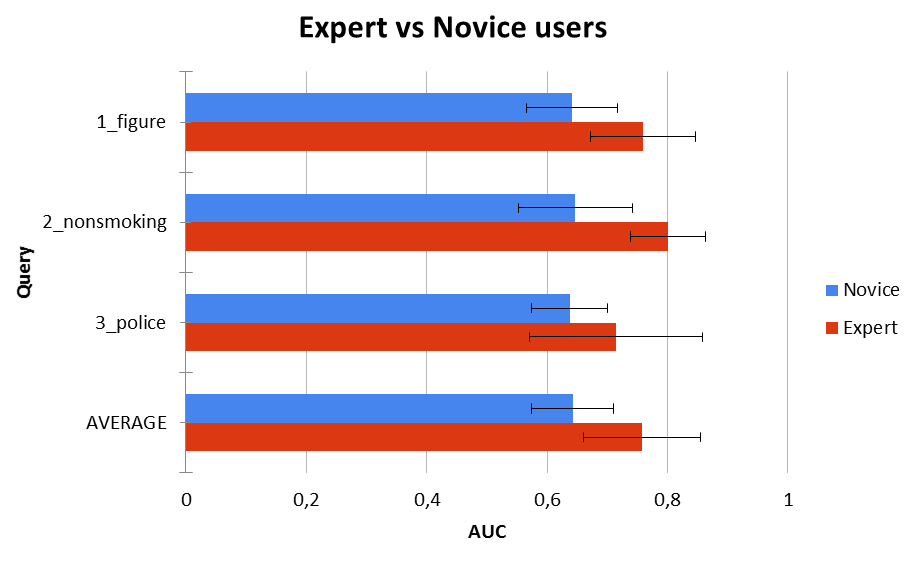}
    \caption[]{Average Area Under the Curve (AUC) for Novice users (blue) and Expert users (red).}
    \label{fig:expertvsnovice}
  \end{center}
\end{figure}

Figure~\ref{fig:expertvsnovice} shows a chart with the average AUC values for users in the two separate groups. We can observe that the average AUC achieved by expert users is significantly higher than the one for novice users (t-test, p=0.005317, sample size = 9). This is a reasonable result, since expert users already know how the images look like prior to the experiment, which gives them an advantage. Another thing that could explain this difference is user motivation: our expert users are involved in the project and want to achieve the best results, so they probably pay more attention to the task than novice users.

\subsection{Visualization rates: 5Hz vs 10Hz}
\label{ssec:visualizationrates}

\begin{figure}[!ht]
  \begin{center}
    \includegraphics[width=\columnwidth]{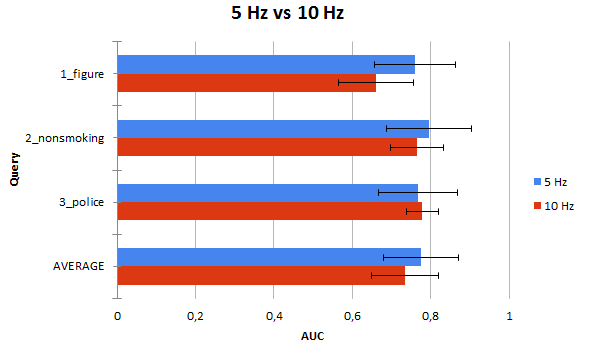}
    \caption[]{Average Area Under the Curve (AUC) for users who did the experiment at 5Hz (blue) and at 10Hz (red).}
    \label{fig:10vs5}
  \end{center}
\end{figure}

The second experiment explores the impact of different visualization rates in users' performance. We compare the performance achieved by those users who did the experiment at 10Hz with those who did it at 5Hz. Figure~\ref{fig:10vs5} shows a chart with the average AUC values for the two configurations. On average, and for our pool of users, the 5Hz visualization rate provides better results. However, the difference between the two is not significant (t-test, p = 0.1397, sample size = 12), which leads us to conclude that the 10Hz is a reasonable visualization rate for users to be able to identify objects in complex images and, more importantly, reduces the length of the experiment by half. This rate comparison is further analyzed in Section~\ref{sec:eegvsmouse} of this paper.

\section{EEG vs. Mouse for Retrieval}
\label{sec:eegvsmouse}

In this section we compare the EEG interaction presented in the previous sections with traditional mouse-based interaction. To do this, we evaluate their performance in a retrieval scenario. First, we use both EEG and mouse annotations to sort the 1000 for each query, generating a ranking of images. Second, we use the EEG and mouse annotations to retrieve relevant images from a larger dataset in a relevance feedback strategy. We use mean Average Precision (mAP) to evaluate the rankings in both cases.

\subsection{The mouse setup}
\begin{figure}[!ht]
  \begin{center}
    \includegraphics[width=\columnwidth]{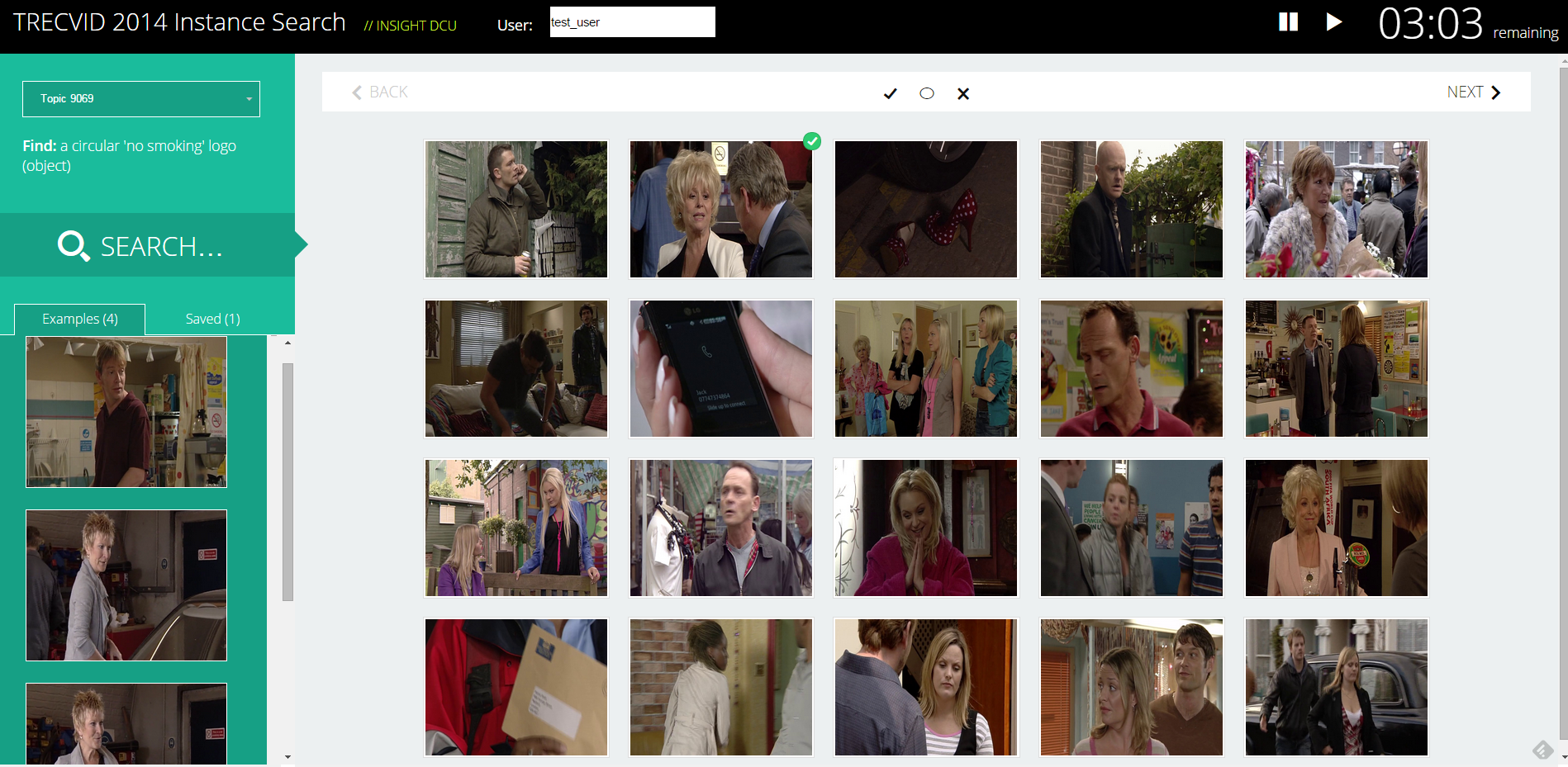}
    \caption[]{The mouse-based user interface.}
    \label{fig:mouseui}
  \end{center}
\end{figure}

Figure~\ref{fig:mouseui} shows a screenshot of the mouse-based user interface that was used for this experiment. Similarly to the EEG setup, this interface displays, for each query: (a) its textual description (b) the four visual examples containing the query. Once the user clicks the \emph{Search} button, the timer starts and the images are displayed. In this setup, users are asked to go through as many results as they can (by clicking the \emph{Next} button) and left-click on the relevant images they find until time is up. 

To make a fair comparison of the mouse and EEG based setups, two considerations have been taken into account:

\begin{itemize}

\item The time given to the users to annotate the images has been restricted to the same amount of time that users spend visualizing images in the EEG setup, i.e. 200 seconds for the 5Hz configuration and 100 seconds for the 10 Hz one.

\item The order in which the images are displayed in the mouse-based interface is the same as the one in the EEG setup (5 blocks of 200 images presented one after the other, keeping a 5\% target ratio in each one of the blocks).

\end{itemize}

We collected annotations from 6 users, this time using the mouse-based interface setting the timer at 200 seconds. To compare with their results, we use the EEG data collected on the 6 users that did the experiment at a visualization rate of 5Hz. We also take the data from the users who did the EEG experiment at 10Hz, to compare with the 100 seconds mouse-based interaction.

\subsection{Retrieval within our dataset}
\label{ss:ranlkin_lists}

In the first experiment, we use the annotations obtained with both EEG and mouse-based mechanisms to sort the 1000 images for each query and produce a ranking.  

The ranking for the mouse setup is constructed as follows: Given a set of positive annotations and their time stamps, we define two sets of images $p_a$ and $n_a$, where $p_a$ are the positive annotations themselves and $n_a$ contains all those unmarked images presented before the last positive annotation (i.e. we assume that all the observed images that have not been clicked are negative). Then, the ranking is built to ensure that the images in $p_a$ are always on top and the images in $n_a$ are always at the bottom. The remaining set of images are placed in between in the same order in which they were displayed in the mouse interface. 

The ranking for EEG is constructed by sorting all the images by their SVM classifier score in descending order.

\subsubsection{Results}
\label{ss:5_results}
\begin{figure}[!ht]
  \begin{center}
    \includegraphics[width=\columnwidth]{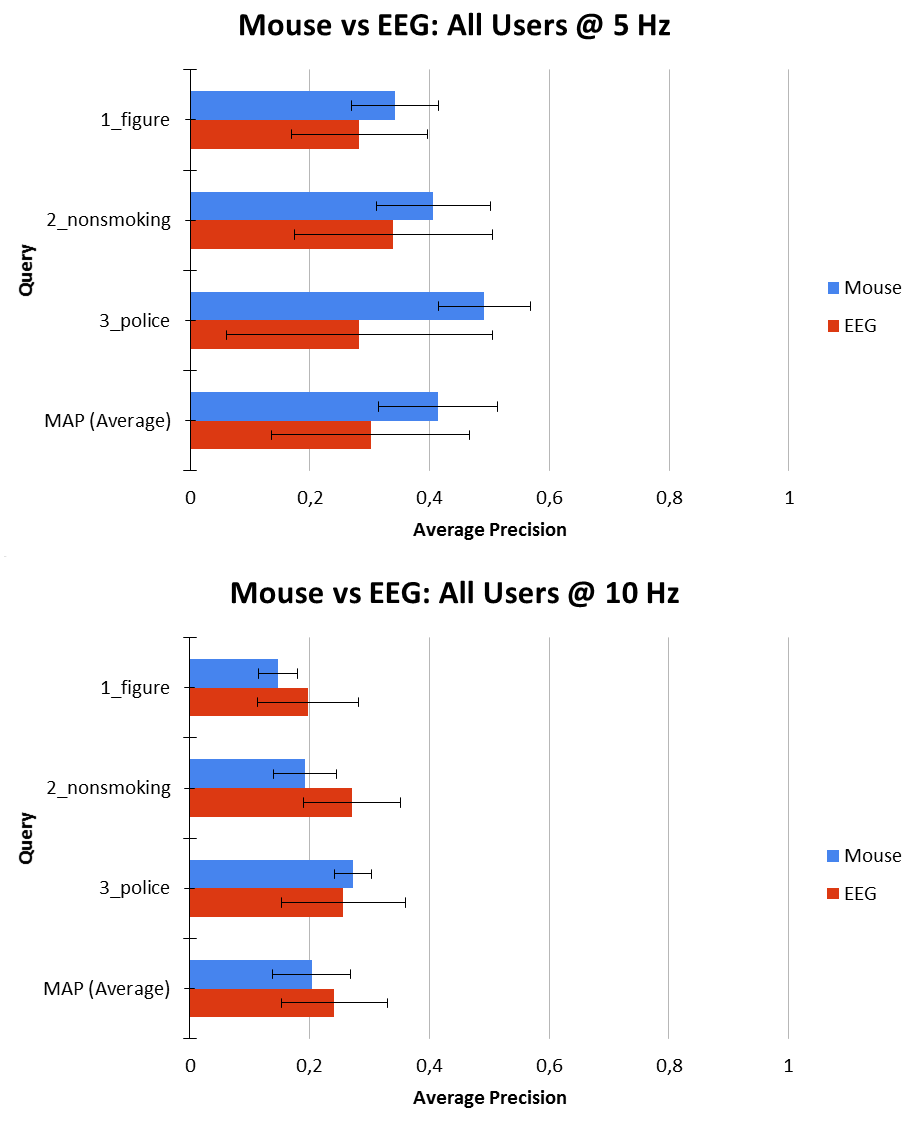}
    \caption[]{Mouse vs EEG. On top, averaged values for all Users at 5Hz / 200 seconds. On the bottom, for 10Hz / 100 seconds.}
    \label{fig:5hz10hz}
  \end{center}
\end{figure}

Figure~\ref{fig:5hz10hz} shows the average performance per user in the three different queries in the two configurations: 5Hz / 200 seconds (top) and 10Hz / 100 seconds (bottom). Several conclusions can be drawn from these results. 


First, there is an accuracy drop in both Mouse and EEG when the annotation time is reduced from 200 to 100. Nevertheless, the drop in performance is a lot higher for the Mouse-based mechanism. The same time limitation on the EEG approach does not seem to affect so much in the final accuracy, which points out the potential of EEG signals, specially when the images are displayed at high frequency rates. 

Second, we can see that the comparison between Mouse and EEG reaches a different conclusion depending on the duration of the experiment. For the 5Hz / 200 seconds experiment, the mAP values obtained for the Mouse are higher than the ones reached with the EEG approach. Conversely, for the 10Hz / 100 seconds experiment, the EEG configuration is better than the Mouse one. 

However, as mentioned in earlier sections, the diversity between users who performed the same experiment is high, which makes the comparison between the two approaches difficult.

\subsection{Retrieval in a larger dataset}
\label{sec:RelevanceFeedback}

In this section we combine the collected annotations with visual descriptors to retrieve more images related to the queries in a bigger dataset. The aim is to compare the relevance feedback provided by the annotations from the mouse and BCI interfaces in a more realistic scenario.

The chosen dataset consists of 23614 frames selected from the TRECVid Instance Search dataset from 2013. We extract visual descriptors for all our images from a Convolutional Neural Network (CNN) pre-trained with ImageNet models \cite{ILSVRCarxiv14} using the \textit{Caffe} software \cite{jia2014caffe}. The selected feature vector is the output of Layer 7 (the second top fully connected layer), which has 4096 dimensions.

We train a linear SVM model with default parameters for each query from the obtained annotations using the CNN descriptors associated with the images. This model is used to score to all images in the larger dataset with scores to the distance of the CNN descriptor to the learned hyperplane. The images are sorted according to their score to generate the final ranking.

Training the models requires positive and negative examples for each query. These examples are generated from the  annotations as follows:

\begin{itemize}

\item For the mouse-based annotations, we use the same approach as in Section~\ref{ss:ranlkin_lists}: the positive annotations are the ones that the user has clicked and the negatives are all those that the user has seen and not clicked.

\item  For the EEG annotations, the annotations are not binary but confidence scores given by the EEG models. Since these annotations are noisy, we only take a small percentage of them to train the model. First, we sort the images according to their EEG score in descending order and we take the first 10 images as positive and the last 100 as negative examples. This criterion has been applied for all the users to simplify the process, even though not all users have the same performance and the percentage of annotations that we trust should vary accordingly.
\end{itemize}

\subsubsection{Results}
\label{ssec:6_results}

Based on the results of the Section~\ref{ss:ranlkin_lists}, when the user interaction time is limited to 100s (10Hz RSVP), click-based and BCI interfaces are comparable in terms of performance.

\begin{figure}[!ht]
  \begin{center}
    \includegraphics[width=\columnwidth] {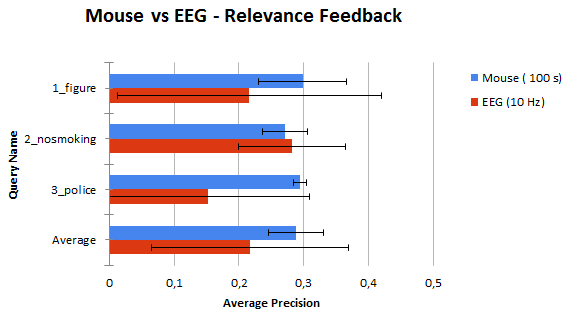}\caption[]{Average Precision comparison between mouse and EEG annotations}
    \label{fig:RF_MouseEEG}
  \end{center}
\end{figure}

Figure~\ref{fig:RF_MouseEEG} shows the performance on the the big dataset for this configuration. On average, the AP obtained using the mouse annotations is higher than the one using EEG. However, these results are averaged on all the users, and we know that there is a high variability among them (in this figure, it is depicted with the standard deviation) and that this variability is especially present in the EEG case. This makes the comparison between the two mechanisms difficult.
 
\begin{figure}[!ht]
  \begin{center}
   \includegraphics[width=\columnwidth]{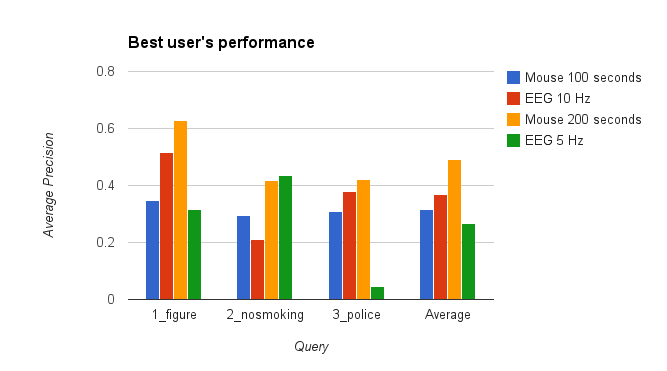}
    \caption[]{Average Precision comparison between mouse and EEG annotations for the best user in both configurations}
    \label{fig:bestUser}
  \end{center}
\end{figure}

To make a comparison that is not subject to user diversity,  Figure~\ref{fig:bestUser} shows the performance of the best user for both interfaces, in the two time configurations. As expected, when user interaction time is set to 200s (5Hz RSVP) the mouse-based annotations are more effective than the EEG ones, obtaining mean AP of 0.49 vs 0.27. Nevertheless, when we consider the best users for the the 100s configuration (10Hz RSVP) we obtain a similar performance for the two, being EEG a bit better on average with a mAP value of 0.37 against a 0.32 for mouse-based feedback.

\section{Discussion}
\label{sec:discussion}

In this paper, we have presented and studied the potential application of EEG as a mechanism towards relevance feedback and compared it to the traditional ``click-based'' one for an object retrieval task. 

We have collected brain signals from 8 different users, and have demonstrated that it is indeed possible to identify objects in images with these types of annotations. The results that we have obtained suggest that users' reactions to relevant images are not only triggered by the object of interest itself, but also by the context in which the user has seen the object in the past (i.e. in the four example images that are presented at the beginning of the the experiment). 

The performance achieved using these brain signals varies a lot from one user to another. Such diversity could be caused by many factors (attention level, fatigue, noise, etc.) that are not considered in this paper. However, we have tried to assess the importance of users' familiarity with the images. We have shown that the accuracy achieved by those users who have been previously exposed to the images is significantly higher than that of those who see them for the first time.

Two different RSVP rates have been tested in our experiments: 5Hz and 10Hz. The comparison between the two has led us to conclude that despite that the performance of the 10Hz configuration is slightly worse, it is also twice as fast. This means that displaying the images at a 10Hz visualization rate allows to annotate the images in less time (compared to 5Hz) with a small drop in accuracy.

We have compared our EEG-based annotation mechanism to a mouse-based one in a retrieval scenario and we have shown that EEG can be as effective as a mouse, for the same amount of time of user interaction. This is specially true in the case when the duration of the experiment is set to 100 seconds (RSVP 10Hz). Increasing the time to 200 seconds (RSVP 5Hz) significantly favors the mouse-based mechanism. What is actually interesting is that, when the interaction time is reduced by two, the drop in performance for the click-based annotation is a lot higher than for the EEG-based one.

The annotations collected with both EEG and mouse interfaces were used to retrieve relevant images from a larger subset of TRECVid images. We showed that EEG annotations can indeed be used in a relevance feedback scenario to retrieve relevant images that have never been presented to the user. Again, we showed that EEG can be as effective as a mouse for these type of tasks, although this is not always true. Mouse annotations are binary and clean, therefore they can generally be trusted. EEG annotations, on the other hand, are noisy and, even with strong filtering, are more likely to contain false positive annotations.

Future work will aim at using this EEG-based system to solve an interactive run of TRECVid through relevance feedback. We foresee a system that will automatically select sets of images to be shown in sequential iterations of RSVP rounds. The relevance feedback captured in each round will be used to refine the final rank of images, as well as to choose which images to be included in the following iteration. Such a system will requires determining the amount of images to include in each round, which should be long enough to capture valid EEG data from a RSVP presentation, but not too long to allow the retrieval system to rapidly adjust to users' feedback. In addition, the sorting of images in each round must be carefully chosen to facilitate the detection of P300 signals, which is most clearly triggered when a target image is presented surrounded by distractors.

\section*{Acknowledgements}

This research was supported by Science Foundation Ireland SFI/12/RC/2289.
This work has been developed in the framework of the project BigGraph TEC2013-43935-R, funded by the Spanish Ministerio de Economía y Competitividad and the European Regional Development Fund (ERDF).
The Image Processing Group at the UPC is a SGR14 Consolidated Research Group recognized and sponsored by the Catalan Government (Generalitat de Catalunya) through its  AGAUR office.
We gratefully acknowledge the support of NVIDIA Corporation with the donation of the GeoForce GTX Titan Z used in this work.

%
\bibliographystyle{abbrv}
\bibliography{sigproc}  
%
%

\end{document}